




\font\bigbold=cmbx12
\font\eightrm=cmr8
\font\sixrm=cmr6
\font\fiverm=cmr5
\font\eightbf=cmbx8
\font\sixbf=cmbx6
\font\fivebf=cmbx5
\font\eighti=cmmi8  \skewchar\eighti='177
\font\sixi=cmmi6    \skewchar\sixi='177
\font\fivei=cmmi5
\font\eightsy=cmsy8 \skewchar\eightsy='60
\font\sixsy=cmsy6   \skewchar\sixsy='60
\font\fivesy=cmsy5
\font\eightit=cmti8
\font\eightsl=cmsl8
\font\eighttt=cmtt8
\font\tensmc=cmcsc10


\def\eightpoint{%
\textfont0=\eightrm   \scriptfont0=\sixrm
\scriptscriptfont0=\fiverm  \def\rm{\fam0\eightrm}%
\textfont1=\eighti   \scriptfont1=\sixi
\scriptscriptfont1=\fivei  \def\oldstyle{\fam1\eighti}%
\textfont2=\eightsy   \scriptfont2=\sixsy
\scriptscriptfont2=\fivesy
\textfont\itfam=\eightit  \def\it{\fam\itfam\eightit}%
\textfont\slfam=\eightsl  \def\sl{\fam\slfam\eightsl}%
\textfont\ttfam=\eighttt  \def\tt{\fam\ttfam\eighttt}%
\textfont\bffam=\eightbf   \scriptfont\bffam=\sixbf
\scriptscriptfont\bffam=\fivebf  \def\bf{\fam\bffam\eightbf}%
\abovedisplayskip=9pt plus 2pt minus 6pt
\belowdisplayskip=\abovedisplayskip
\abovedisplayshortskip=0pt plus 2pt
\belowdisplayshortskip=5pt plus2pt minus 3pt
\smallskipamount=2pt plus 1pt minus 1pt
\medskipamount=4pt plus 2pt minus 2pt
\bigskipamount=9pt plus4pt minus 4pt
\setbox\strutbox=\hbox{\vrule height 7pt depth 2pt width 0pt}%
\normalbaselineskip=9pt \normalbaselines
\rm}


\def\pagewidth#1{\hsize= #1}
\def\pageheight#1{\vsize= #1}
\def\hcorrection#1{\advance\hoffset by #1}
\def\vcorrection#1{\advance\voffset by #1}

\newcount\notenumber  \notenumber=1              
\newif\iftitlepage   \titlepagetrue              
\newtoks\titlepagefoot     \titlepagefoot={\hfil}
\newtoks\otherpagesfoot    \otherpagesfoot={\hfil\tenrm\folio\hfil}
\footline={\iftitlepage\the\titlepagefoot\global\titlepagefalse
           \else\the\otherpagesfoot\fi}

\def\abstract#1{{\parindent=30pt\narrower\noindent\eightpoint\openup
2pt #1\par}}
\def\smc{\tensmc}


\def\note#1{\unskip\footnote{$^{\the\notenumber}$}
{\eightpoint\openup 1pt
#1}\global\advance\notenumber by 1}

\def\frac#1#2{{#1\over#2}}

\def\({\left(}
\def\){\right)}
\def\<{\langle}
\def\>{\rangle}
\def\2pd#1#2#3{\frac{\partial^2#1}{\partial#2\partial#3}}

\def\sqr#1#2{{\vcenter{\vbox{\hrule height.#2pt
        \hbox{\vrule width.#2pt height#1pt \kern#1pt
           \vrule width.#2pt}
        \hrule height.#2pt}}}}
\def\square{\mathchoice\sqr64\sqr64\sqr{4.2}3\sqr33}
\def\ni{\noindent}
\def\ref #1{$^{[#1]}$}
\def\slash{\!\!\!\!/}
\def\lqq{\lq\lq}
\def\rqq{\rq\rq}


\def\mpole{{M}}
\def\d{{\rm d}}

\def\ps{\rlap{{\it p}}\kern0.4pt\lower2pt\hbox{/} \hskip3pt}


\pageheight{24cm}
\pagewidth{15.5cm}
\hcorrection{-2.5mm}
\magnification \magstep1
\baselineskip=19pt
\parskip=5pt plus 1pt minus 1pt


\rightline {MZ-TH/94-09}
\vskip 50pt
\centerline{\bigbold The Nielsen Identities for the}
\centerline{\bigbold Two-Point Functions of QED and QCD}
\vskip 27pt
\centerline{\smc J.C. Breckenridge$^{*}$, M.J. Lavelle{\hbox
{$^{\dag}$}}{\note{e-mail: lavelle@vipmza.physik.uni-mainz.de}}
and  T.G. Steele{\hbox {$^{*}$}}{\note{e-mail: steelet@sask.usask.ca}}}
\vskip 15pt
{\baselineskip 12pt
\centerline{\null$^{*}$Department of Physics and Engineering Physics}
\centerline{ and Saskatchewan Accelerator Lab}
\centerline{University of Saskatchewan}
\centerline{Saskatoon}
\centerline{Saskatchewan, S7N 0W0}
\centerline{Canada}
\vskip 12pt
\centerline{\null$^{\dag}$Institut f\"ur Physik}
\centerline{Johannes Gutenberg-Universit\"at}
\centerline{Staudingerweg 7, Postfach 3980}
\centerline{D-55099 Mainz, F.R.\thinspace Germany}
}
\vskip0.5truecm
\centerline{{\sl To appear Z. Phys. C}}
\vfill\eject
\vglue 1truecm
{\baselineskip=13pt\parindent=0.58in\narrower\ni{\bf Abstract}\quad
We consider  the Nielsen  identities  for the two-point functions of full
QCD and QED in the class of
Lorentz  gauges.   For pedagogical  reasons  the identities are first derived
in QED to demonstrate the gauge independence of the photon self-energy,
and of the
electron mass shell.
In QCD we derive the general  identity  and hence
the identities  for the quark, gluon and ghost propagators.
The explicit contributions to the gluon and ghost identities are
calculated to one-loop order,
and then we show that the quark identity requires that in on-shell
schemes the
quark mass renormalisation must be gauge independent. Furthermore,
we obtain formal solutions
for the gluon self-energy and ghost propagator in terms of the gauge
dependence of other, independent Green functions.}

\ni{\bf PACS No.:}\quad 11.30.Ly \quad 12.38.Aw\quad 12.20.Ds
\par

\vfill\eject

\ni {\bf 1. Introduction}

\ni The Nielsen  identities\ref{1}  deserve  to be better  known.
They  follow  from  a modification  of  Becchi-Rouet-Stora-Tyutin
(BRST) symmetry in a similar way to how the usual Ward identities
are obtained from BRST.  The identities  are, however,  of a quite
different nature to the Ward identities.  This may be illustrated
by      (very)      schematically       writing      them      as
$\frac{\partial}{\partial\xi}G_1=G_2  G_3$,  where  $\xi$  is the
gauge parameter and the $G_i$ are Green's functions.   This then,
for  example,  clearly  tells  us that  if the  right  hand  side
vanishes,  then $G_1$ must be gauge  independent.   The aim of this
paper  is  to  investigate   the  consequences   of  the  Nielsen
identities for the two-point functions of QED and QCD.

We will not follow here the original derivation of Nielsen, but rather
use an alternative approach due to Piguet and Sibold\ref{2}.  The
identities  have  been previously  employed\ref{3}  to study  the
gauge  invariance  of the  Higgs  mass  and the  r\^ole  of gauge
symmetry  in the effective  potential  for models  of spontaneous
symmetry breaking.

It will become evident in what follows that the Nielsen identities are
particularly useful for investigations of on-mass shell Green's functions
and on-shell renormalisation constants. The on-shell renormalisation scheme
is widely used in QED, the electroweak theory and QCD with heavy quarks and
it is of the greatest importance for studies of the S-matrix.

For completeness  and clarity we begin by discussing  the case of
QED in Section  2.  We show in particular that the pole mass of the
electron  is gauge  independent  and that  the photon  self-energy
can  be
simply  shown  to  be  gauge  parameter
independent.  We stress here the interplay between the usual (BRST) Ward
identities and the Nielsen identities.

In Section 3 we present the
Nielsen identities for QCD.
We explicitly calculate the one-loop contributions  in the identity
for the gluon propagator, and then demonstrate
that in on-shell renormalisation schemes
the quark mass renormalisation must
also be gauge  independent  to all orders.  The Nielsen identity for the ghost
propagator is constructed and
its explicit content is determined to one-loop order.
Formal solutions for the gluon self-energy and ghost propagator
appear from this analysis which deserve further study.

Section 4 provides a perturbative analysis of the Nielsen identities
illustrating the
gauge parameter independence of the mass renormalisation in the mass-shell
scheme. Some conclusions are presented in Section 5.

\bigskip
\ni{\bf 2. The QED Identities}

The standard path integral formulation of QED in covariant gauges is based
upon the following Lagrangian
$$ {\cal L}_{\eightpoint QED} = -\frac14 F^2 + \bar\psi(i D\!\slash -m) \psi
 +\frac{\xi}2 B^2+ B \partial\cdot A - \bar c\,\square\, c \,,
\eqno (2.1)
$$
\ni where $\xi$ is the gauge parameter, $F$ is the field strength
and $B$ is an auxiliary field\ref{4}.
Although ghosts decouple in QED, one often includes them so as to demonstrate
that the Lagrangian is invariant under the following BRST transformations
$$
 \eqalign{\delta A_\mu & =\epsilon \partial_\mu c\,,\cr
\delta B & =0\,,}\quad\eqalign{
\delta \bar\psi & = +i\epsilon c \bar\psi\,,\cr
\delta \psi  &=-i\epsilon c\psi\,,}\quad\eqalign{
\delta c &=0\,,\cr
\delta \bar c & =\epsilon B\,.\cr}
\eqno (2.2)
$$
The Ward identities may then be obtained by exploiting this invariance in the
usual fashion\ref{5}.

Following [2], the Nielsen identities are obtained by making the following
addition to the Lagrangian
$$ {\cal L}_{\eightpoint QED}\rightarrow {\cal L}_{\eightpoint QED} +
\frac\chi 2 \bar c B \,,
\eqno (2.3)
$$
\ni where $\chi$ is a global Grassmannian variable, $\chi^2=0$. (In the
following care must be taken with the minus signs that are needed under the
interchange of Grassmannian variables!) It is clear
upon a little reflection
that the addition of this term cannot change the dynamics of the theory.
To see this most easily consider that the generating functional can be
expanded in $\chi$ and, as a result of its Grassmannian nature, only two
terms survive:
$\chi^0$ and $\chi^1$. The first term gives us the usual dynamics of QED
and the second must vanish by virtue of ghost number
when we integrate over the ghost fields. This
shows that we have not changed any physics. In what follows we will employ
the Lagrangian $(2.3)$.

This modified Lagrangian is invariant under the following extended set of
BRST transformations
$$ \eqalign{\delta^+ A_\mu & =\epsilon \partial_\mu c\,,\cr
\delta^+ B & =0\,,\cr
\delta^+ \psi & =-i\epsilon c\psi\,,\cr
\delta^+ \bar\psi & = +i\epsilon c \bar\psi\,,\cr}\qquad\eqalign{
\delta^+ \bar c & =\epsilon B\,,\cr
\delta^+ c &=0\,,\cr
\delta^+ \xi & =\epsilon \chi\,,\cr
\delta^+ \chi & =0\,.\cr
}
\eqno (2.4)
$$
\ni where $\epsilon$ is a Grassmann quantity.
The $F^2$ and Dirac Lagrangians are, as usual, separately invariant
and the
remaining terms are readily seen to be invariant. The crucial point
to note is that the gauge parameter is now transformed.

To exploit this invariance and so derive the Nielsen identities we now
consider the following generating functional
$$
\eqalign{Z= \int \![d\mu]\, \exp\bigg\{
i\int\!\!d^4x\, {\cal L}_{\eightpoint QED}  + J_\mu A^\mu + &
\bar J_\psi \psi + \bar\psi J_{\bar\psi} + B J_{B}  \cr
&  + \bar K_\psi (-ic\psi) + ic \bar\psi K_{\bar\psi}
\bigg\}\,.\cr }
\eqno (2.5)
$$
The various sources denoted by $J$ with a subscript are standard ones. We
have not included sources for the ghost fields, for example, since we
will not consider their Green's functions, for QED they are anyway
trivial. The purpose of the additional, rather exotic
looking, sources denoted by
$K$'s will become apparent in a moment. Note that we may rewrite this part
of the Lagrangian as
$$
\bar K_\psi\frac{\delta^+ \psi}{\delta\epsilon} +
\frac{\delta^+ \bar\psi}{\delta\epsilon} K_{\bar\psi}
 \,.
\eqno (2.6)
$$

To study the gauge dependence of the electron and photon propagators,
we now introduce
the generating functional of proper Green functions.
$$
\eqalign{\Gamma(A^\mu, \psi, \bar\psi, & B, c, \bar c, \chi, \xi,
  \bar K_\psi, K_{\bar\psi})
= \cr & W(J_\mu,
J_{\bar\psi}, \bar J_\psi, J_B, K_{\bar\psi}, \bar K_\psi, \chi, \xi)
-\int\!\!d^4x\, J_\mu A^\mu +J_BB +\bar J_\psi \psi + \bar\psi J_{\bar\psi}\cr}
\eqno (2.7)
$$
As a consequence of invariance under (2.4) we have
$$
\delta^+\Gamma \equiv 0=\delta^+A_\mu \frac{\delta \Gamma}{\delta A_\mu}
+\delta^+\psi \frac{\delta \Gamma}{\delta \psi}
+\delta^+\bar\psi \frac{\delta \Gamma}{\delta \bar\psi}
+\delta^+\bar c \frac{\delta \Gamma}{\delta \bar c}
+\delta^+\xi \frac{\partial\Gamma}{\partial\xi}
\,,
\eqno(2.8)
$$
where we have used $\delta^+B=\delta^+c=\delta^+\chi=0$.
{}From (2.6) we have
$\delta^+\psi=\epsilon\frac{\delta\Gamma}{\delta\bar K_\psi}$
etc. We may therefore rewrite (2.8) as
$$
0=\partial_\mu c \frac{\delta \Gamma}{\delta A_\mu}
+ \frac{\delta\Gamma}{\delta \bar K_\psi} \frac{\delta \Gamma}{\delta \psi}
+  \frac{\delta\Gamma}{\delta K_{\bar\psi}}
\frac{\delta \Gamma}{\delta \bar\psi}
+  B\frac{\delta \Gamma}{\delta \bar c}
+\chi\frac{\partial\Gamma}{\partial\xi}
\,.
\eqno(2.9)
$$
Note that implicit coordinate space integrations are understood in the above!

We are now in a position to obtain the Nielsen identities for QED. We
differentiate (2.9) with respect to $\chi$ and then set $\chi$ to zero.
$$
\eqalign{
0&=\frac{\partial\Gamma}{\partial\xi}
- \partial_\mu c \frac{\delta^2\Gamma}{\delta\chi\delta A_\mu}
+ \frac{\delta^2\Gamma}{\delta\chi\delta\bar{K}_\psi}
\frac{\delta\Gamma}{\delta\psi}
-\frac{\delta\Gamma}{\delta \bar{K}_\psi}
\frac{\delta^2\Gamma}{\delta\psi\delta\chi}
\cr
&+\frac{\delta^2\Gamma}{\delta\chi\delta
K_{\bar\psi}}\frac{\delta\Gamma}{\delta{\bar\psi}}
+\frac{\delta\Gamma}{\delta
K_{\bar\psi}}\frac{\delta^2\Gamma}{\delta\chi\delta{\bar\psi}}
+B\frac{\delta^2\Gamma}{\delta\chi\delta\bar{c}}
}
\eqno(2.10a)
$$
In the special case when no further functional derivatives with respect
to ghost fields
will be applied to (2.10a) ghost number
conservation implies a simplification.
The so simplified result is then

$$
0=\frac{\partial\Gamma}{\partial\xi}
+B\frac{\delta^2\Gamma}{\delta\chi\delta{\bar c}}
+\frac{\delta^2\Gamma}{\delta\chi\delta{\bar K}_\psi}
\frac{\delta\Gamma}{\delta\psi}
+\frac{\delta^2\Gamma}{\delta\chi\delta K_{\bar \psi}}
\frac{\delta\Gamma}{\delta{\bar \psi}}
\, ,
\eqno (2.10b)
$$
where we have used the result
$\partial _\mu c=\partial_\mu \frac{\delta \Gamma}{\delta J_{\bar c}}=0$
for the one-point function.
{}From these central results we can generate the QED Nielsen identities
for the two-point functions.

\bigskip
\ni {\bf 2.1 The Mass of the Electron}

The first application studied here is the electron mass. Although the
the fermion field $\psi$ is not
BRST-invariant and may not be naively identified with the
electron\ref{6,7},
we will now show that its pole mass is gauge
independent and may so be given a physical meaning.
The inverse fermion propagator is given by
$iS^{-1}
(y-x)=\frac{\delta^2\Gamma}{\delta\psi(y)\delta\bar\psi(x)}$.
Differentiating (2.10b) with respect to $\psi(y)$ and ${\bar\psi}(x)$
we obtain
$$
\frac{\partial}{\partial\xi}\frac{\delta^2\Gamma}{\delta\psi(y)\delta
\bar\psi(x)}  =
+
\frac{\delta^3\Gamma}{\delta\psi(y)\delta\bar K_\psi\delta \chi}
\frac{\delta^2\Gamma}{\delta\bar\psi(x)\delta\psi}
+
\frac{\delta^2\Gamma}{\delta\psi(y)\delta\bar\psi}
\frac{\delta^3\Gamma}{\delta\bar\psi(x)\delta K_{\bar\psi}\delta\chi}
\quad ,
\eqno (2.11)
$$
and all other terms will vanish, either by fermion conservation or by their
relation
to one-point functions.

To complete the analysis of the Fermion propagator (2.11) must be transformed
into
momentum space.  Defining
$$
\eqalign{
\frac{\delta^2\Gamma}{\delta\chi\delta {\bar K}_\psi(w)\delta\psi (y)}
&=\int \frac{d^4q}{(2\pi)^4}\frac{d^4\ell}{(2\pi)^4}\,e^{-iq\cdot(y-z)
-i\ell\cdot(w-z)}
F(q,\ell,-q-\ell)\cr
\frac{\delta^2\Gamma}{\delta\chi\delta K_{\bar\psi}(w)\delta{\bar\psi} (y)}
&=\int \frac{d^4q}{(2\pi)^4}\frac{d^4\ell}{(2\pi)^4}\,e^{-iq\cdot(x-z)
-i\ell\cdot(w-z)}
{\bar F}(q,\ell,-q-\ell) \, ,
}
\eqno (2.12)
$$
and recalling that (2.11) contains implicit $w$ and $z$ integrations,
it is found that
$$
\frac{\partial}{\partial\xi}S^{-1}(p)=S^{-1}(p)\bigl\{ F(p,-p,0)+
\bar F(-p,p,0) \bigr\}
\, .
\eqno(2.13)
$$
This result is the Nielsen identity for the inverse fermion propagator. In
particular, since the right hand side vanishes at the mass shell,
and since $F(-p,p,0)$ has no single particle poles, we have
$$
\left.\frac{\partial S^{-1}(p)}{\partial\xi}\right|_{p^2={M}^2}=0
\,,
\eqno(2.14)
$$
where ${M}$ is the pole mass.

{}From this result it is easy to see that $\mpole$ is gauge
independent. Since we may, in a covariant gauge, generally write the
inverse propagator as
$$
S^{-1}(p)=A(p^2)p\slash - B(p^2)
\,,
\eqno(2.15)
$$
then $\mpole$ is defined by
$$
A(\mpole^2) \mpole=B(\mpole^2)
\,.
\eqno (2.16)
$$
If we differentiate this equation with respect to $\xi$ and compare it with
(2.14) rewritten in the following way
$$
\left.\left(\frac{\partial A(p^2)}{\partial\xi} p\slash-
\frac{\partial B(p^2)}{\partial\xi}\right)\right|_{p^2=\mpole^2}=
\mpole \frac{\partial A(\mpole^2)}{\partial\xi} -
\frac{\partial B(\mpole^2)}{\partial\xi}
=0
\,,
\eqno (2.17)
$$
we see that we obtain the desired result
$$
\left[A+\frac{\partial A}{\partial \mpole}-\frac{\partial B}{\partial \mpole}
\right]\,
\frac{\partial \mpole}{\partial \xi}=\left[
\frac{\partial B(\mpole^2)}{\partial\xi}-
\mpole \frac{\partial A(\mpole^2)}{\partial\xi} \right]
=0 \,\rightarrow\, \frac{\partial \mpole}{\partial \xi}=0
\eqno(2.18)
$$
so that  the pole  mass  is gauge  parameter independent
(note that the bracketed quantity on the left hand side of (2.18) is non-zero).
In Section 4 this point will be
studied more carefully in a perturbative fashion.

\bigskip
\ni {\bf 2.2 The Photon Propagator}

\ni The inverse photon propagator may be studied in a similar fashion to
the above. By functional differentiation of (2.10b) with respect to
$A_\nu(x)$ and $A_\lambda(y)$, we find
$$
\frac{\partial}{\partial\xi}\frac{\delta^2\Gamma}{\delta A_\nu(x)\delta
A_\lambda(y)}=0
\,, \eqno  (2.19)
$$
where  many terms  that must,  from considerations  of fermion and ghost
number or relation to one-point functions, vanish have been neglected.
At this stage it is important to recognize that the auxiliary field $B$ is
independent
of $A_\mu$.  To make a direct connection between (2.19) and the photon vacuum
polarization,
it is necessary to consider some aspects of the auxiliary field formalism
\ref{4,8}.

To formulate perturbation theory it is necessary to consider the free field
case,
corresponding to the quadratic part of the Lagrangian (2.1).  The mixing
between
$B$ and $\partial\cdot A$ in the Lagrangian must be diagonalized after
functional integration,
leading to the following {\sl free field} bosonic propagators.
$$
\eqalignno{
\int d^4x\,e^{ip\cdot x}\langle O\vert T\left(B(x) B(0)\right)\vert O\rangle
&=0 &(2.20a)\cr
\int d^4x\,e^{ip\cdot x}\langle O\vert T\left(B(x) A_\mu(0)\right)\vert O
\rangle &= \frac{p^\mu}{p^2} &(2.20b)\cr
\int d^4x\,e^{ip\cdot x}\langle O\vert T\left(A_\mu(x) A_\nu(0)\right) \vert O
\rangle &=
i\left[-\frac{g^{\mu\nu}}{p^2} +(1-\xi)\frac{p^\mu p^\nu}{p^4}\right] &(2.20c)
}
$$

BRS symmetry implies that (2.20a) and (2.20b) are true to all orders in
perturbation theory.
$$
\eqalignno{
0 &\!=\!\frac{\delta_{BRS}}{\delta\epsilon}\,\,\langle O\vert T
\left(\bar c(x)B(y)\right)\vert O\rangle =
\langle O\vert T\left(B(x) B(y)\right)\vert O\rangle &(2.21a)\cr
0 &\!=\!\frac{\delta_{BRS}}{\delta\epsilon}\!\langle O\vert T\!
\left(\bar c(x)A_\mu(y)\right)\!\vert O\rangle =
\langle O\vert T\!\left(B(x) A_\mu(y)\right)\!\vert O\rangle\! +\!
\langle O\vert T\!\left(\partial_\mu c(x) \bar c(y)\right)\!\vert O\rangle
&(2.21b)
}
$$
By recognizing that the ghosts are decoupled, (2.21b) then implies that
(2.20b) is valid to all
orders in perturbation theory.  This then implies that radiative corrections
to the tree-level
result (2.20b) are absent, so the photon vacuum polarization  is necessarily
transverse
as illustrated in Figure 1.
$$
0=p^\mu \Pi_{\mu\lambda}\left( -\frac{g^{\nu\lambda}}{p^2}+ (1-\xi)
\frac{p^\nu p^\lambda}{p^4}\right)
\eqno(2.22)
$$
Hence we have the usual result for the full photon propagator.
$$
\int d^4x\,e^{ip\cdot (x-y)}\langle O\vert T\left(A_\mu(x) A_\nu(y)\right)
\vert O\rangle
=i\left[ -\frac{g^{\mu\nu}}{p^2}+\frac{p^\mu p^\nu}{p^4}\right] \,
\frac{1}{1+\Pi(p^2)}
-i\xi\frac{p^\mu p^\nu}{p^4}
\eqno (2.23)
$$

The implications of the results for the Green functions (2.20a), (2.20b) and
(2.23) for the
generating functional $\Gamma$ can be understood by consideration of the
following functional identities.
$$
\eqalign{
\frac{\delta B(x)}{\delta A_\mu(y)} &=0=\frac{\delta A_\mu(x)}{\delta B(y)}\cr
\frac{\delta B(x)}{\delta B(y)} &= \delta(x-y)\cr
\frac{\delta A_\mu(x)}{\delta A_\nu (y)}&=\delta(x-y)\,\delta_\nu^\mu
}
\eqno(2.24)
$$
In terms of generating functionals these expressions become
$$
\eqalign{
\delta(x-y)\delta_\nu^\mu &=\int d^4z
\frac{\delta^2\Gamma}{\delta A_\lambda(z)\delta A_\nu(y)}
\frac{\delta^2W}{\delta J_\lambda(z)\delta J_\mu(x)}
+\frac{\delta^2\Gamma}{\delta B(z)\delta A_\nu(y)}
\frac{\delta^2W}{\delta J_B(z)\delta J_\mu(x)}  \cr
0 &=\int d^4z
\frac{\delta^2\Gamma}{\delta A_\lambda(z)\delta B(y)}
\frac{\delta^2W}{\delta J_\lambda(z)\delta J_\mu(x)}
+\frac{\delta^2\Gamma}{\delta B(z)\delta B(y)}
\frac{\delta^2W}{\delta J_B(z)\delta J_\mu(x)}  \cr
0 &=\int d^4z
\frac{\delta^2\Gamma}{\delta A_\lambda(z)\delta A_\mu(y)}
\frac{\delta^2W}{\delta J_\lambda(z)\delta J_B(x)}
+\frac{\delta^2\Gamma}{\delta B(z)\delta A_\mu(y)}
\frac{\delta^2W}{\delta J_B(z)\delta J_B(x)}  \cr
\delta(x-y) &=\int d^4z
\frac{\delta^2\Gamma}{\delta A_\lambda(z)\delta B(y)}
\frac{\delta^2W}{\delta J_\lambda(z)\delta J_B(x)}
+\frac{\delta^2\Gamma}{\delta B(z)\delta B(y)}
\frac{\delta^2W}{\delta J_B(z)\delta J_B(x)}
}
\eqno(2.25)
$$
The functional derivatives of $W$ are known from the Green functions
discussed above, and hence (2.25) implies that the quadratic pieces of $\Gamma$
simply diagonalize the various $A_\mu$, $B$ two-point Green functions in
momentum space.
$$
\eqalignno{
\int d^4x\,e^{ip\cdot x}\,\frac{\delta^2\Gamma}{\delta A_\mu(x)\delta A_\nu(0)}
&= p^2\left[ -g_{\mu\nu}+p_\mu p_\nu\right]\,\left(1+\Pi(p^2)\right) &(2.26a)
\cr
\int d^4x\,e^{ip\cdot x}\,\frac{\delta^2\Gamma}{\delta A_\mu(x)\delta B(0)}
&=p_\mu &(2.26b) \cr
\int d^4x\,e^{ip\cdot x}\,\frac{\delta^2\Gamma}{\delta B(x)\delta B(0)}
&=\xi &(2.26c)
}
$$

The transversality of (2.26a) is easily seen to be consistent with the
fundamental identity (2.9).
\footnote\dag{We are grateful to the referee for bringing this point to
our attention.}
Upon functional differentiation of (2.9) with respect to $A_\mu$ and $c$,
setting
$\xi=0$, and then Fourier transforming, we see that
$$
p^\mu\,\int d^4x\,e^{ip\cdot x}\,\frac{\delta^2\Gamma}{\delta A_\mu(x)
\delta A_\nu(0)} = 0
\eqno(2.27)
$$
consistent with the analysis leading to (2.26a).

Finally, substituting the result of (2.26a) into the identity (2.19)
we find that the polarization  $\Pi(p^2)$ must be independent of
the gauge parameter\ref{7}
$$ \frac{\partial \Pi(p^2)}{\partial\xi} \equiv 0
\,.
\eqno(2.24)
$$
This result shows the power of the Nielsen  identities.
\bigskip
Since the ghosts decouple this concludes our survey of the
Nielsen identities for the two-point functions of QED.

\bigskip
\ni {\bf 3. The QCD Identities}

\ni The reader who has closely followed the above will have no difficulty
in finding the analogous identities for QCD\ref{9}. We give the
basic steps and so define our notation. (Note that colour indices are
implicit.) The modified QCD Lagrangian is
$$
{\cal L}_{\eightpoint QCD} = -\frac14 F^2 + \bar\psi(i D\!\slash -m) \psi
 +\frac{\xi}2 B^2+ B \partial\cdot A - \bar c\,\partial^\mu D_\mu\,c \,
 + {\chi \over 2}\bar c B
\eqno (3.1)
$$
which is invariant under the following augmented BRST transformations
$$
 \eqalign{\delta^+ A_\mu & =\epsilon D_\mu c\,,\cr
\delta^+ B & =0\,,\cr
\delta^+ \psi & =-i\epsilon c\psi\,,\cr
\delta^+ \bar\psi & = +i\epsilon c \bar\psi\,,\cr}\qquad\eqalign{
\delta^+ \bar c & =\epsilon B\,,\cr
\delta^+ c &=-{1 \over 2} \epsilon [\, c\, , c]\,,\cr
\delta^+ \xi & =\epsilon \chi\,,\cr
\delta^+ \chi & =0\,.\cr
}\eqno (3.2)
$$
As in the QED case, the extension of BRST includes a transformation of
the gauge parameter.

The generating functional~$Z$ with sources (including ghosts) is
$$
\eqalign{Z= \int \![d\mu]\, \exp\bigg\{
i\int\!\!d^4x\, &{\cal L}_{\eightpoint QCD}  + J_\mu A^\mu + J_BB
+\bar J_\psi \psi + \bar\psi J_{\bar\psi} + \bar{J}_c c+
\bar{c} J_{\bar c}\cr
&+ K_\mu (D^\mu c)
-\frac{1}{2} \bar K_{c} [\, c\, , c]
 + \bar K_\psi (-ic\psi) + ic \bar\psi K_{\bar\psi}
\bigg\}\,.\cr }
\eqno (3.3)
$$
and the generating functional~$\Gamma$ for the proper Green functions is
$$
\eqalign{\Gamma(A^\mu, \psi, \bar\psi,  B, c, \bar c, \chi, \xi,
 K_\mu,  \bar K_\psi, K_{\bar\psi})
=  & W(J_\mu,J_B,
J_{\bar\psi}, \bar J_\psi, \bar{J}_c , J_{\bar c},
K_\mu, K_{\bar\psi}, \bar K_\psi, \chi, \xi) \cr
&\,-\int\!\!d^4x\, J_\mu A^\mu +J_B B + \bar J_\psi \psi + \bar\psi
J_{\bar\psi} +\bar{J}_c c + \bar{c} J_{\bar c} \cr}
\eqno (3.4)
$$
{}From this last result it is simple to obtain the nonabelian version of (2.9).
$$
0=\frac{\delta\Gamma}{\delta K_\mu}\frac{\delta\Gamma}{\delta A_\mu}
+\frac{\delta\Gamma}{\delta {\bar K}_\psi}\frac{\delta\Gamma}{\delta\psi}
+\frac{\delta\Gamma}{\delta K_{\bar \psi}}\frac{\delta\Gamma}{\delta\bar\psi}
+B\frac{\delta\Gamma}{\delta\bar c}
+\frac{\delta\Gamma}{\delta {\bar K}_c}\frac{\delta\Gamma}{\delta c}
+\chi\frac{\partial \Gamma}{\partial\xi}
\eqno (3.5a)
$$
After differentiation with respect to $\chi$ and then setting $\chi=0$ we find
$$
\eqalign{
0&=\frac{\partial\Gamma}{\partial\xi}
+ \frac{\delta^2\Gamma}{\delta\chi\delta K_\mu}
\frac{\delta\Gamma}{\delta A_\mu}
- \frac{\delta\Gamma}{\delta K_\mu}\frac{\delta^2\Gamma}{\delta\chi\delta
A_\mu}
+ \frac{\delta^2\Gamma}{\delta\chi\delta\bar{K}_\psi}
\frac{\delta\Gamma}{ \delta\psi}
\cr
&-\frac{\delta\Gamma}{\delta \bar{K}_\psi}
\frac{\delta^2\Gamma}{\delta\psi\delta\chi}
+\frac{\delta^2\Gamma}{\delta\chi\delta K_{\bar\psi}}
\frac{\delta\Gamma}{\delta{\bar\psi}}
+\frac{\delta\Gamma}{\delta K_{\bar\psi}}
\frac{\delta^2\Gamma}{\delta\chi\delta{\bar\psi}}
+B\frac{\delta^2\Gamma}{\delta\chi\delta \bar c}
\cr
&+\frac{\delta^2\Gamma}{\delta\chi\delta\bar K_c}\frac{\delta\Gamma}{\delta c}
+\frac{\delta\Gamma}{\delta\bar{K}_c}\frac{\delta^2\Gamma}{\delta\chi\delta c}
}
\eqno(3.5b)
$$
which  is the QCD equivalent  of (2.10a).  As before, further functional
differentiation
with respect to fundamental non-ghost fields can be obtained from the following
simpler result using ghost number conservation in (3.5b).
$$
0=\frac{\partial\Gamma}{\partial\xi}
+ \frac{\delta^2\Gamma}{\delta\chi\delta K_\mu}
\frac{\delta\Gamma}{\delta A_\mu}
+ \frac{\delta^2\Gamma}{\delta\chi\delta\bar{K}_\psi}
\frac{\delta\Gamma}{ \delta\psi}
+\frac{\delta^2\Gamma}{\delta\chi\delta K_{\bar\psi}}
\frac{\delta\Gamma}{\delta{\bar\psi}}
+B\frac{\delta^2\Gamma}{\delta\chi\delta \bar c}
\eqno(3.5c)
$$

Although expressions (3.5c) and (2.10b) are superficially very similar, the
distinction between abelian and non-abelian theories shows up in
two important ways.  First, there is now the current $K_\mu$ which couples
ghosts and gauge fields
through the covariant derivative
and secondly the ghosts are no longer decoupled.
\bigskip
\ni {\bf 3.1 Nielsen Identity for the Gluon Propagator}

\ni  The distinction between non-abelian and abelian theories is most
evident when considering the Nielsen identity for the gauge field
propagators. To our knowledge, these identities have not been
previously studied in an explicit calculation.
Proceeding as in Section 2.2, one finds
$$
\frac{\partial}{\partial\xi}
\frac{\delta^2\Gamma}{\delta A_\nu(x)\delta A_\lambda(y)}
=\frac{\delta^2\Gamma}{\delta K_\mu\delta\chi\delta A_\nu(x)}
\frac{\delta^2\Gamma}{\delta A_\mu\delta A_\lambda(y)}
+ \nu\leftrightarrow\lambda\quad .
\eqno(3.6a)
$$
Defining the Green functions,
$$
\eqalignno{
 F_ {\mu \nu} (p, q,-p-q) &= \int d^4y\int\!\!\d^4x \,e^{i p \cdot x+iq\cdot y}
  {{\delta^3\Gamma}\over{\delta A_
 \mu (x) \delta K_ \nu (y)\delta \chi}} \,.
 &(3.6b)\cr
\Gamma_{\mu\nu}(p) &=\int d^4x\,e^{ip\cdot (x-y)}
\frac{\delta^2\Gamma}{\delta A_\mu(x)\delta A_\mu(y)}
&(3.6c)
}
$$
we have the following result after Fourier transforming (3.5a).
$$
\frac{\partial}{\partial \xi}\Gamma_{\nu\lambda}(p)
=\Gamma_{\lambda\mu}(p)F_{\mu\nu}(-p,p,0)+ \nu\leftrightarrow \lambda
\eqno(3.6d)
$$

As in the photon analysis, it is necessary to relate the gluon vacuum
polarization to
$\Gamma_{\mu\nu}(p)$ before studying (3.6d).  The analysis is only slightly
more complicated
than in the abelian case.  The free field results (2.20) clearly are
identical, but
the BRS symmetry is somewhat different.
$$
\eqalignno{
0 &=\frac{\delta_{BRS}}{\delta\epsilon} \langle O\vert T\left(\bar c(x) B(y)
\right)\vert O\rangle
=\langle O\vert T\left( B(x) B(y)\right) \vert O\rangle & (3.7a)\cr
0 &=\frac{\delta_{BRS}}{\delta\epsilon} \langle O\vert T
\left(\bar c(x) A_\mu (y)\right)\!\vert O\rangle
\!=\!\langle O\vert T\left( B(x) A_\mu(y)\right)\!\vert O\rangle
\!-\!\langle O \vert T\left( \bar c(x) D_\mu c(y)\right)\!\vert O\rangle
&(3.7b)
}
$$
The relevant Green function in (3.7b) is
$$
\int d^4x\,e^{i p\cdot x}\langle O\vert T\left(\bar c(0) D_\mu c(x)\right)
\vert O\rangle
=\frac{p^\mu}{p^2}
\eqno (3.7c)
$$
as can be verified by contracting both sides with $p^\mu$ then applying the
ghost
equation of motion and the canonical commutation relations.

Thus we see from (3.7) that the abelian results of (2.20a) and (2.20b)
apply to all orders in QCD.
The same argument as outlined in Section 2.2 then implies that the gluon
vacuum polarization
must be transverse, and so (2.26a) is also valid in QCD.

In contrast to the case of QED, (3.6d) cannot be solved  exactly since
the Green function $F_{\mu\nu}(p,-p,0)$ must be determined
perturbatively. However, we can explicitly display the perturbative
contributions in this identity, specifically to one-loop order.
Since the vacuum polarization $\Pi(p^2)$ is well known (see, e.g., [8])
we only have to calculate  $F_{\mu\nu}(p,-p,0)$ .

When calculating $F_{\mu\nu}(p,-p,0)$
it is important to recognize that the mixing between the fields $B$ and $A_\mu$
implies that we do not simply have a truncated Green function.  This can
be seen in the
following way:
$$
\eqalign{
\frac{\delta^3\Gamma}{\delta A_\mu(x)\delta K_\nu(y)\delta \chi}
&=\int d^4 s\,
\frac{\delta^3 W}{\delta K_\nu(y)\delta\chi\delta J_\lambda(s)}
\frac{\delta J_\lambda(s)}{\delta A_\mu(x)}
+\frac{\delta^3 W}{\delta K_\nu(y)\delta\chi\delta J_B(s)}
\frac{\delta J_B(s)}{\delta A_\mu(x)}\cr
&=\int d^4s
\frac{\delta^3 W}{\delta K_\nu(y)\delta\chi\delta J_\lambda(s)}
\frac{\delta^2\Gamma}{\delta A_\mu(x)\delta A_\lambda (s)}\cr
&\quad + \int d^4 s
\frac{\delta^3 W}{\delta K_\nu(y)\delta\chi\delta J_B(s)}
\frac{\delta^2\Gamma}{\delta A_\mu(x)\delta B(s)}
}
\eqno(3.8a)
$$
When the relevant Fourier transform is taken and the result of (2.26b)
is applied,
the second term becomes
$$
\eqalign{
\frac{1}{2}p^\mu &\int d^4xd^4y\,e^{ip\cdot x-i p\cdot y}\langle O\vert
T\left(D_\nu c(y) B(x) \bar c(0) B(0)\right)\!\vert O\rangle\cr
&\quad=\frac{1}{2}\int d^4x\, e^{ip\cdot x}\langle O\vert T
\left( B(x) B(0)\right)\!\vert O\rangle =0
}
\eqno(3.8b)
$$
where the ghost equation of motion, commutation relations, and
(2.20a) have been applied.

Thus after Fourier transforming (3.8a) we have
$$
F_{\mu\nu}(p,-p,0)={\cal F}_{\mu\lambda}(p,-p,0)\,\Gamma_{\lambda\nu}(p)
\eqno(3.8c)
$$
where
$$
{\cal F}_{\nu\lambda}(p,-p,0)=\frac{1}{2}\int d^4xd^4y \,e^{ip\cdot x-ip\cdot
y}
\langle O\vert T\left(D_\nu c(y) A_\lambda (x) \bar c(0)B(0)\right)\!
\vert O\rangle
\eqno(3.8d)
$$
is the full Green function.

To  one loop order, the contributions to this Green function are given
by the diagrams of Fig.\thinspace 2.
The results of the calculation for the contribution to ${\cal F}_{\mu\nu}$
from individual diagrams of Fig.\thinspace 2 are
$$
\eqalign{
2{\cal F}_{\mu \nu}^{(\romannumeral 1)} =& -\frac{g^2N_c}{64\pi^2p^4}
p_\mu p_\nu\left[ -\frac{3\xi}{\hat \epsilon} + 4\xi\right]\cr
2{\cal F}_{\mu \nu}^{(\romannumeral 2)}=& \frac{g^2N_c}{64\pi^2p^4}
\left[ p^2g_{\mu\nu}\left( -2\xi-\frac{3}{\hat\epsilon}\right)
+p_\mu p_\nu\left(-\frac{3\xi}{\hat\epsilon}+6\xi+
\frac{3}{\hat \epsilon}\right)\right]\cr
2{\cal F}_{\mu \nu}^{(\romannumeral 3)}=&\frac{g^2N_c}{64\pi^2p^4}
\left[p^2g_{\mu\nu}\left(\frac{1}{\hat\epsilon}-2\right)
+p_\mu p_\nu\left(\frac{\xi}{\hat\epsilon}-
\frac{1}{\hat\epsilon}+2\right)\right]\cr
2{\cal F}_{\mu \nu}^{(\romannumeral 4)}=&\frac{g^2N_c}{64\pi^2p^4}
p_\mu p_\nu\left[\frac{\xi}{\hat\epsilon}-\frac{3}{\hat\epsilon}+4\right]\cr
2{\cal F}_{\mu \nu}^{(\romannumeral 5)}=& -\frac{g^2N_c}{64\pi^2p^4}
p_\mu p_\nu\left[\frac{\xi}{\hat\epsilon}-\frac{3}{\hat\epsilon}+4\right]\cr
2{\cal F}_{\mu \nu}^{(\romannumeral 6)}=&-\frac{g^2N_c}{64\pi^2p^4}
p_\mu p_\nu\left[\frac{\xi}{\hat\epsilon}\right]\cr
}
\eqno(3.9a)
$$
The sum of these diagrams, including the tree level contribution is
$$
2{\cal F}_{\mu \nu}=\frac{p^\mu p^\nu}{p^4}  -{{g^2 N_c} \over{64 \pi^2
p^4}} \left(p^2 g_{\mu \nu} - p_ \mu p_ \nu \right) \left({2 \over {\hat
\epsilon}} + 2 + 2 \xi \right)\,  .\eqno (3.9b)
$$
Note that we have introduced
$$
\frac{1}{\hat \epsilon}=\frac{1}{\epsilon}-\log{4\pi}+\gamma +
\log{\left(-\frac{p^2}{\nu^2}\right)}\qquad D=4+2\epsilon \,,
\eqno(3.9c)
$$
to be consistent with\ref{10}, and included a factor of $2$ to take care of
the ``crossed'' term in (3.6d).

It is important that this Green function is explicitly transverse
beyond tree-level as required by conservation of the ghost current
$K_\mu=D_\mu c$, which follows from ordinary BRST symmetry.
This is easily seen by considering the Green function ${\cal F}_{\mu\nu}$.
$$
\eqalign{
\frac{1}{2}q^\nu\int d^4x d^4y \,&e^{ip\cdot x+iq\cdot y}\langle O\vert
T\left(A_\mu(x) D_\nu c(y) \bar c(0) B(0)\right)\!\vert O\rangle\cr
&=\frac{1}{2}\!\int d^4 x e^{ip\cdot x} \langle O\vert T
\left( A_\mu(x) B(0)\right)\!\vert O\rangle\! =\! \frac{p^\mu}{2p^2} \,,
}
\eqno(3.9d)
$$
where as usual, the ghost equations of motion and commutation relations
have been applied.
Clearly the identity (3.9d) is satisfied by the tree level contribution
$p^\mu q^\nu/(2p^2q^2)$, so the higher-loop contributions to
${\cal F}_{\mu\nu}(p,-p,0)$
must be transverse, serving as a consistency check on our calculation.

Returning to the identity (3.6d) and including terms up to one-loop we have
the following result for $\Gamma_{\lambda\mu}(p)$ \ref{10}
$$
\Gamma_{\lambda \mu} (p) = \left(g_{\mu \nu} p^2 - p_ \mu p_ \nu \right)
\left(1 + \Pi(p^2) \right)
$$
where
$$
\Pi (p^2) =  {{g^2 N_c} \over{16 \pi^2}} \left[ \left({{13} \over{6}} -
{\xi \over 2} \right) {1 \over{\hat \epsilon}} - {{97} \over{36}}
- {1 \over 2} \xi - {{\xi^2} \over{4}} \right]\,,
\eqno(3.10a)
$$
and some terms independent of the gauge parameter have been neglected.
These neglected terms in (3.10a) come from quarks, and are irrelevant for
the analysis of (3.6d).
Combining (3.10a) with the results of (3.9) to one-loop order yields
$$
2F_{\mu\nu}(p,-p,0)=  -{{g^2 N_c} \over{64 \pi^2
p^2}} \left(p^2 g_{\mu \nu} - p_ \mu p_ \nu \right) \left({2 \over {\hat
\epsilon}} + 2 + 2 \xi \right)\,
\eqno (3.10b)
$$
which confirms the QCD Nielsen identity (3.6d) for the gluon propagator.

A formal solution for the gluon self-energy $\Pi(p^2)$ can be obtained
from (3.6) and
from the transverse nature of $F_{\mu\nu}(p,-p,0)$. Defining
$$
2F_{\mu\nu}(p,-p,0)=\left( g_{\mu\nu} - \frac{p^\mu
p^\nu}{p^2}\right) F^{(1)}(p^2)
\eqno(3.11a)
$$
then using (3.10) and (3.6) we have
$$
\eqalign{
\frac{\partial\Pi(p^2)}{\partial\xi}&=\left[1+\Pi(p^2)\right] F^{(1)}(p^2)\cr
\frac{\partial}{\partial\xi}\log\left[1+\Pi(p^2)\right]&= F^{(1)}(p^2)\cr
1+\Pi(p^2)&=\exp\left[\int  F^{(1)}(p^2) d\xi\right]
}
\eqno(3.11b)
$$
This formal solution relating the gluon self-energy to another Green
function could be
of interest in other contexts, but at present we do not see any immediate
implications.
However, this result could provide a different approach for studying
questions of confinement.

\vfill\eject
\ni {\bf 3.2 Gauge Dependence of the Quark Propagator}

The Nielsen identity for the {\it quark} propagator can be obtained
directly from (3.5b) by functional
differentiation with respect to~$\psi(x)$, $\overline{\psi}(y)$ and
using quark number conservation.
$$
\eqalign{\frac{\partial}{\partial\xi}\frac{\delta^2\Gamma}{\delta\psi(y)
\delta
\bar\psi(x)} & =  \frac{\delta^2\Gamma}{\delta K_\mu\delta\chi}
\frac{\delta^3\Gamma}{
\delta \psi(y)\delta\bar\psi(x) \delta A^\mu}
+
\frac{\delta^4\Gamma}{ \delta \psi(y)\delta\bar\psi(x) \delta K_\mu\delta\chi}
\frac{\delta \Gamma}{\delta A^\mu}\cr
& +
\frac{\delta^3\Gamma}{\delta\psi(y)\delta\bar K_\psi\delta\chi}
\frac{\delta^2\Gamma}{\delta\bar\psi(x)\delta\psi}
+
\frac{\delta^2\Gamma}{\delta\psi(y)\delta\bar\psi}
\frac{\delta^3\Gamma}{\delta\bar\psi(x)\delta K_{\bar\psi}\delta\chi}
\cr} \,.
\eqno (3.12)
$$
As in Section 2, the first and second terms are zero because of
Lorentz invariance with operator insertions at zero momentum.  This
leads to a result identical to (2.13)\ref{9}.
$$
\frac{\partial}{\partial\xi}S^{-1}(p)=S^{-1}(p)\left\{ F(p,-p,0)+
\bar F(p,-p,0) \right\}\,.
\eqno(3.13)
$$
Again, since $S^{-1}(p)$ is zero at the mass shell, and since
$F(p,-p,0) $ has no single particle pole, we have the following Nielsen
identity for the quark propagator.
$$
{{\partial S^{-1}} \over{\partial \xi}} (p) \bigg\vert_{p^2 = M^2} = 0 \,.
\eqno(3.14)
$$
Just as in Section 2, this result may be shown to imply that
the quark mass shell is gauge
independent in QCD for on-shell renormalisation schemes.
We will return to a perturbative consideration of
this idea in Section 4.

\bigskip
\ni{\bf 3.3 Nielsen Identity for the Ghost Propagator}

To construct the Nielsen identity for the ghost propagator we need to
differentiate~(3.5a) with respect to~$ c(x)$ and~$\bar c(0)$.  After
applying ghost
and fermion  number conservation the following identity is obtained:
$$
\eqalign{
{\partial \over{\partial \xi}} {{\delta^2 \Gamma} \over{\delta c(x)
\delta \bar c(0)}} = & -{{\delta^2 \Gamma} \over{\delta \chi \delta
K_ \mu (w)}}{{\delta^3 \Gamma} \over{\delta c(x) \delta \bar c(0)
\delta A_ \mu (w)}} -
{{ \delta^2 \Gamma} \over{\delta K_ \mu(w) \delta c(x)}}{{ \delta^3 \Gamma}
\over{\delta \chi \delta \bar c(0) \delta A_ \mu(w)}}\cr
& +{{\delta^2 \Gamma} \over{\delta c(w) \delta\bar c(0)}}
{{ \delta^3 \Gamma} \over{\delta \chi \delta \bar K_c (w) \delta c(x)}}
+ B(w)\frac{\delta^4\Gamma}{\delta\chi\delta\bar c(w)\delta c(x)\delta\bar
c(0)}
\,.
}
\eqno{(3.15)}
 $$

The dependence on the variable~$w$ (which has an integration associated with
it) has been  made explicit to avoid possible confusion with~$c(x)$.
The first term on the right hand side of (3.15) is zero since after Fourier
transforming we have a
zero momentum insertion with a vector operator.  The third term is also zero
since the one point function has the property $B=\delta W/\delta J_B=0$.

To analyze the remaining terms we first recognize that
$$
\int d^4(x-y)\,e^{ip\cdot (x-y)}
\frac{\delta^2\Gamma}{\delta c(x)\delta K_\mu (y)} =
\frac{p^\mu}{p^2} \tilde D^{-1}(p^2)
\eqno(3.16)
$$
where $\tilde D^{-1}(p^2)$ is the inverse ghost propagator.  After Fourier
transforming
(3.15) we have the following Nielsen identity for the ghost propagator.
$$
\eqalignno{
\frac{\partial}{\partial\xi}\tilde D^{-1}(p^2)&=\frac{p^\mu}{p^2}
G_\mu(p,-p,0)\tilde
D^{-1}(p^2)+G^{(1)}(p,-p,0)\tilde D^{-1}(p^2)
&(3.17a)\cr
G^{(1)}(p,q,-p-q)&\equiv\int d^4xd^4y\,e^{ip\cdot x+i q\cdot y}
\frac{\delta^3\Gamma}{\delta \bar K_{c}(x)\delta c(y)\delta\chi}
&(3.17b)\cr
G^\mu(p,q,-p-q)&\equiv \int d^4xd^4y\,e^{ip\cdot x+i q\cdot y}
\frac{\delta^3\Gamma}{\delta A_\mu(x)\delta\bar c(y)\delta\chi}
&(3.17c)
}
$$
Now consider the Green function $G^\mu$ to take into account the mixing between
$A_\mu$ and $B$.
$$
\eqalign{
\frac{\delta^3\Gamma}{\delta\chi\delta\bar c(y)\delta A_\mu(x)}
&=\int d^4sd^4w \frac{\delta^3 W}{\delta \chi\delta \bar J_c (w)J_\lambda (s)}
\frac{\delta \bar J_c(w)}{\delta \bar c(y)}
\frac{\delta J_\lambda(s)}{\delta A_\mu(x)}\cr
&\,+\int d^4s d^4w\frac{\delta^3W}{\delta\chi\delta \bar J_c(w)\delta J_B(s)}
\frac{\delta \bar J_c(w)}{\delta \bar c(y)}\frac{\delta J_B(s)}{\delta
A_\mu(x)}
\cr
&=\int d^4sd^4w \frac{\delta^3 W}{\delta \chi\delta \bar J_c (w)J_\lambda (s)}
\frac{\delta^2\Gamma}{\delta\bar c(y)\delta c(w)}
\frac{\delta^2\Gamma}{\delta A_\mu(x)\delta A_\lambda (s)}\cr
&\,+\int d^4sd^4w
\frac{\delta^3W}{\delta\chi\delta \bar J_c(w)\delta J_B(s)}
\frac{\delta^2\Gamma}{\delta \bar c(y)\delta c(w)}
\frac{\delta^2\Gamma}{\delta A_\mu(x)\delta B(s)}
}
\eqno(3.18)
$$
After Fourier transforming we find that the first term in (3.18) does not
contribute to
$p^\mu G_\mu(p,-p,0)$ since $\Gamma_{\mu\lambda}(p)$ is transverse.
Thus we have
$$
\frac{p^\mu}{p^2}G_\mu(p,-p,0)=\tilde D^{-1}(p^2)\frac{1}{2}\int d^4xd^4y \,
e^{ip\cdot x-ip\cdot y}
\langle O\vert T\left(\bar c(0) B(0) c(x)B(y)\right)\vert O\rangle
\eqno(3.19)
$$
where (2.26b) has been used.

Again, we can explicitly display the perturbative content of (3.17a) to
one-loop by
calculating the diagrams in Figures 3 \& 4 and noting that the tree level
contribution in each case is zero.
The results are
$$
\eqalign{
\frac{p^\mu}{p^2}G_\mu(p,-p,0)&=-\frac{g^2N_c}{64\pi^2}\left[
\frac{1}{\hat \epsilon}-2\right]\cr
G^{(1)}(p,-p,0)&=-\frac{g^2N_c}{64\pi^2}\left[-\frac{2}{\hat\epsilon}+2\right]
}
\eqno(3.20)
$$
The result for the ghost propagator is \ref{10}
$$
\eqalign{
\tilde D^{-1}(p^2)&=p^2\left[ 1+\Pi(p^2)\right]\cr
\Pi(p^2)&=-\frac{g^2N_c}{16\pi^2}\left[\,\frac{3-\xi}{4\hat\epsilon} - 1\right]
}
\eqno(3.21)
$$
Substituting (3.20) and (3.21) into (3.17a) verifies the Nielsen identity
for the ghost propagator to one-loop.

Similarly to the case for the gluon propagator,~(3.17a) has a formal solution.
Defining
$$
G^{(2)}(p,-p,0)=G^{(1)}(p,-p,0)+\frac{p^\mu}{p^2}G_\mu(p,-p,0)
\eqno(3.22)
$$
and noting that there is no dependence on $\tilde D(p^2)$ in (3.22),
we have the
following formal solution to the Nielsen identity.
$$
\tilde D^{-1} (p^2) = p^2\, \exp \left[ \int\!\!\d \xi G^{(2)} (p, -p,0)
\right].
\eqno(3.23)
$$
\bigskip
\ni {\bf 4. On-Shell Fermion Mass Renormalisation Constant}

\ni In this section we wish to investigate the consequences of the Nielsen
identities [see (2.14) and (3.14)] for the fermion propagator in a perturbative
analysis of on-shell renormalisation constants. The key result from both QED
and
QCD is
$$
{{\partial S^{-1}_F } \over{\partial \xi}}(p) \Big\vert_{p^2 = M^2} = 0\,.
\eqno\hbox{(4.1)}
$$

Consider the following definition of the fermion mass
renormalisation constant (valid for both QED and QCD) in an on-shell
scheme\ref{11,12}
$$
Z_m \equiv  {m_0 \over M} \,,\eqno\hbox{{(4.2)}}
$$
where $\ps =\mpole$ is a zero of the inverse propagator.
We will prove that $Z_m$ (and hence $M$) must be gauge parameter independent
to all orders in perturbation theory.

To construct this direct link between the
results of perturbation theory \ref{11-13} and our formal proof of the
gauge independence of the mass-shell we will show that the
coefficients~$M_i$  in the expansion of the mass renormalisation
constant (we are using $D=4-2\omega$ to be consistent with [9])
$$
Z_m \equiv {m_0 \over M} = 1 + \sum_{i=1}^\infty \left(\alpha_s \over
{M^{2 \omega}} \right)^i\, M_i\,, \eqno\hbox{(4.3)}
$$
are gauge independent to all orders in perturbation theory.  This has
been explicitly observed
to two loops in perturbation theory \ref{11-13}.

Now, the Feynman propagator is written, using the conventions of \ref{11}, as
$$
\eqalignno{
S^{-1}_F (p) &= \ps - m_0 - \Sigma (p)\,,
&(4.4)\cr
\Sigma (p) &= \sum_{i=1}^\infty \left( \alpha_s \over{p^{2 \omega}}
\right)^i \left( m_0 A_i \left({m_0^2} \over{p^2} \right) + (\ps - m_0)
B_i\left({m_0^2} \over{p^2} \right)\right)\,.
&(4.5) \cr
}
$$

We must also have, in addition to the above Nielsen identity, that the
inverse propagator itself must vanish on the mass shell~$(p^2=M^2)$.  This is
sufficient to define the coefficients~$M_i$ in the expansion of~$Z_m$ in
the following way.
Working to the one-loop level, we will disregard anything of the order
of~$\alpha_s^2$ or higher.  Thus we have the expression for~$S^{-1}_F (p) $
at one-loop to be
$$
S^{-1}_F (p) = \ps - m_0 - {{\alpha_s} \over{p^{2 \omega}}} \left( m_0 A_1
 \left({m_0^2} \over{p^2} \right) + (\ps - m_0) B_1\left({m_0^2} \over{p^2}
 \right)\right). \eqno\hbox{(4.6)}
$$
We then substitute for~$m_0$ from (4.3), obtaining, to
order~$\alpha_s$
$$
\eqalign{
S^{-1}_F (p) =& \ps-M + M-M\left(1 + {{\alpha_s} \over{M^{2 \omega}}} M_1
\right) \cr &\quad- {{\alpha_s} \over{p^{2 \omega}}}\left( M A_1 \left({M^2}
\over{p^2} \right) + (\ps - M) B_1\left({M^2} \over{p^2} \right)\right). }
\eqno\hbox{(4.7)}
$$
This expression is then evaluated on the mass shell and set equal to
zero:
$$
 S^{-1}_F (p)\Big\vert_{p^2=M^2} = -{{\alpha_s} \over{M^{2 \omega}}} M
 M_1 - {{\alpha_s} \over{M^{2 \omega}}}M A_1 \left(1 \right) = 0\,, \eqno
 \hbox{(4.8)}
$$
thus we must have~$M_1 = -A_1 (1)$ in order for  equation~(4.8) to be
satisfied.

If one now considers the Nielsen identity (4.1), then one
obtains to one loop the relation
$$
{{ \partial S^{-1}_F} \over{\partial \xi}} (p) =  - {{\alpha_s}
 \over{p^{2 \omega}}}\left( m_0 {{ \partial A_1} \over{\partial \xi}}
 \left({m_0^2} \over{p^2} \right) + (\ps - m_0) { {\partial B_1}\over
 {\partial \xi}}\left({m_0^2} \over{p^2} \right)\right)\,.
\eqno\hbox{(4.9)}
$$
Upon going on-shell and substituting for~$m_0$, we get
$$
 {{ \partial S^{-1}_F} \over{\partial \xi}} (p)\Big\vert_{p^2=M^2} =
 - {{\alpha_s} \over{M^{2 \omega}}} M {{ \partial A_1} \over{\partial \xi}}
 \left(1 \right) = 0\,.
\eqno\hbox{(4.10)}
$$
Thus we obtain the result that
$$
{{ \partial A_1} \over{\partial \xi}} \left(1 \right) = 0\,,
\eqno\hbox{(4.11)}
$$
and thus that
$$
{{ \partial M_1} \over{\partial \xi}}  = 0\,, \eqno\hbox{(4.12)}
$$
ie., that~$M_1$ is gauge independent.

It is this result that we wish to argue can be continued to all orders in
perturbation theory.
That this is possible can be seen by comparing the structure of the Nielsen
identity above with that of the determining equation for~$M_1$, {\it before}
one goes
on-shell.
The structure is identical, with the exception of the leading term
containing~$M_1$.
What remains when one  goes on-shell, in (4.7) defines~$M_1$, and what
 remains after going on-shell in (4.9) shows~$M_1$ to be gauge independent.

 Let us illustrate by continuing on to two loops.  To the two-loop level,
 we have
 $$
 \eqalign{
 S^{-1}_F (p) = \ps - m_0 - &{{\alpha_s} \over{p^{2 \omega}}}
 \left( m_0 A_1 \left({m_0^2} \over{p^2} \right) + (\ps - m_0) B_1
 \left({m_0^2} \over{p^2} \right)\right) \cr& -{{\alpha_s^2}
 \over{p^{4 \omega}}} \left( m_0 A_2 \left({m_0^2} \over{p^2} \right)
 + (\ps - m_0) B_2\left({m_0^2} \over{p^2} \right)\right)}\,,
\eqno\hbox{(4.13)}
$$
again we substitute for~$m_0$ and keep all terms of order~$\alpha_s^2$ or
less.  This  gives
$$
 \eqalign{
 S^{-1}_F (p) =& \ps - M - \left({{\alpha_s} \over{M^{2 \omega}}} M
 M_1 + {{\alpha_s^2} \over{M^{4 \omega}}} M M_2 \right) -{{\alpha_s}
 \over{p^{2 \omega}}} \Biggl(M A_1 \left({M^2} \over{p^2} \right)  \cr &\quad+
 {{\alpha_s} \over{M^{2 \omega}}} M M_1 A_1\!\!\left({M^2} \over{p^2} \right)
 + {{\alpha_s} \over{M^{2 \omega}}} {{2 M^3 M_1} \over{p^2}} A_1^\prime\!\!
 \left({M^2} \over{p^2} \right) \cr &\qquad + \left( \ps - M -{{\alpha_s}
 \over{M^{2 \omega}}} M M_1 \right) B_1\left({M^2} \over{p^2} \right)
 \Biggr) \cr &\quad-{{\alpha_s^2} \over{p^{4 \omega}}}\left( M A_2\left({M^2}
 \over{p^2} \right) + (\ps - M) B_2\left({M^2} \over{p^2} \right) \right)\,.}
 \eqno\hbox{(4.14)}
 $$
 Going on-shell, it is easy to see that the terms in~(4.14) of
order~$\alpha_s$ are
simply the one-loop result, and that the order~$\alpha_s^2$
 terms will define $M_2$ to be:
$$
M_2 = -\left(A_2(1) + M_1 \left( A_1(1) + 2 A_1^\prime(1) - B_1(1) \right)
\right)\,. \eqno\hbox{(4.15)}
$$

 On the other hand, we can see that application of the identity~(4.1) to
 the two-loop expression~(4.13) will result in
 $$
 \eqalign{
{{ \partial S^{-1}_F} \over{\partial \xi}} (p) =& -{{\alpha_s}
\over{p^{2 \omega}}} \Biggl(M {{\partial A_1}\over{\partial \xi}}
\left({M^2} \over{p^2} \right) + {{\alpha_s} \over{M^{2 \omega}}}
M {{\partial} \over{\partial \xi}}\left(M_1  A_1\left({M^2} \over{p^2}
\right)\right)  \cr &\qquad+ {{\alpha_s} \over{M^{2 \omega}}} {{2 M^3 }
\over{p^2}} {{\partial }\over{\partial \xi}} \left(M_1 A_1^\prime
\left({M^2} \over{p^2} \right) \right)+ \left( \ps - M \right)
{{\partial B_1}\over{\partial \xi}}\left({M^2} \over{p^2} \right) \cr
&\qquad-{{\alpha_s} \over{M^{2 \omega}}} M {{\partial}\over{\partial \xi}}
\left(M_1   B_1\left({M^2} \over{p^2} \right)\right) \Biggr)\cr &\qquad
-{{\alpha_s^2}
\over{p^{4 \omega}}}\left( M {{\partial A_2}\over{\partial \xi}}\left({M^2}
\over{p^2} \right) + \!(\ps - M) {{\partial B_2}\over{\partial \xi}}
\left({M^2} \over{p^2} \right)\! \right)}\,,\eqno\hbox{(4.16)}
$$
and here we can see that when we go on shell the order~$\alpha_s$ part
reproduces equation~(4.11) and the order~$\alpha_s^2$ part will give
 $$
{{ \partial S^{-1}_F} \over{\partial \xi}} (p) \Big\vert_{p^2=M^2} =
-{{\partial }\over{\partial \xi}} \left(M_1 \left( A_1(1) + 2A_1^\prime(1)
- B_1(1) \right) + A_2(1) \right) = 0\,, \eqno\hbox{(4.17)}
 $$
which obviously guarantees that~$M_2$ is gauge independent.  Note also that
 we made no use of one-loop results in obtaining equations~(4.15) and~(4.17),
 and thus the gauge independence of~$M_2$ in no way depends upon the gauge
 independence of~$M_1$, although the one-loop results will
obviously make things simpler.

One can see, then, that the relationship between the defining equation
for~$M_n$ and the~$n$-loop Nielsen identity~(4.1), which shows the gauge
independence of~$M_n$, will be maintained to all orders in perturbation
theory.

Having demonstrated that $Z_m$ is gauge independent ($\partial Z_m/\partial\xi
=0$)
it then follows from (4.2) that $\partial M/\partial\xi=0$.
$$
\eqalign{
\frac{\partial Z_m}{\partial\xi}+\frac{\partial M}{\partial \xi}
\frac{\partial Z_m}{\partial M}
&= - \frac{Z_m}{M}\frac{\partial M}{\partial\xi}\cr
\frac{\partial M}{\partial\xi}\left[\frac{Z_m}{M}
-\frac{\partial Z_m}{\partial M}\right]
&=0 \quad\rightarrow \quad \frac{\partial M}{\partial \xi}=0
}
\eqno(4.18)
$$
This completes our perturbative analysis of the gauge independence of the
mass shell and mass renormalisation constant in on-shell renormalisation
schemes.

\bigskip
\ni {\bf 5. Conclusions}

\ni In this paper we have studied  the Nielsen identities  for the
two-point functions of QED
and QCD in the covariant formalism.   It was demonstrated in the case of
QED that the identities  lead to results complementary to those
of the usual Ward identities. As with
the Ward  identities, the Nielsen identities offer possibilities to check
one's calculations, however, they also allow us to see where physical
meaning may be found in apparently gauge dependent Green's functions.
In particular  it was proven that
the  electron  pole  mass,  the  photon  polarization  and  the on-shell
mass renormalisation constant $Z_m$
are all independent of the gauge parameter.

For  QCD  it  was  demonstrated  that  the  quark  pole  mass
and on-shell mass renormalisation constant are
gauge independent. This is a formal property and we are not sure as
to its correct interpretation  in relation  to quark confinement.
 In particular the generating function used above to describe QCD,
does not take the Gribov ambiguity\ref{14} into account. It has
recently been shown that this ambiguity prevents the construction
outside of perturbation theory of a
globally BRST invariant field with quark number one\ref{15} which
could be identified with a physical quark.  {\it Perturbatively\/}
invariant quark fields do exist, however, and we stress the need to
clarify the connection between the invariance of the pole mass,
shown here using functional methods, and the perturbatively
gauge invariant solutions with quark number one found in [15].
In this context we note that the gauge independence of the pole
mass has already been shown\ref{13} to hold in perturbation
theory  up to two loops  and appears  to be true  in the operator
product expansion (OPE) of the quark propagator\ref{9,16}, although
care must be taken there since gauge-dependent condensates appear
in the OPE of the gauge-dependent propagator\ref{17}.

We have calculated the explicit one-loop content of the Nielsen identities
obtained for the gluon and ghost propagators.
Formal solutions to the ghost and gluon Nielsen identities were constructed
for the propagators in terms independent Green functions.  Since this
gives a different view of the gluon propagator this solution may be
of interest for studies of confinement.

\bigskip
\ni{\bf Acknowledgements:} MJL thanks D. Broadhurst and K. Schilcher for
discussions, the University of Saskatchewan for their hospitality during a
visit to Saskatoon and the Graduierten Kolleg of the University of Mainz
for support. TGS thanks V. Elias for discussions, and
is grateful for the financial support of the Natural Sciences and
Engineering Research council of Canada.  We are indebted to the referee
for their constructive comments on the misuse of the auxiliary field
formalism in
the original version of the manuscript.

\vfill\eject
\ni {\bf References}

\itemitem{1.)}{N.K.\ Nielsen, Nucl.\ Phys.\ {\bf B101} (1975) 173.}
\itemitem{2.)}{O.\ Piguet \& K.\ Sibold, Nucl.\ Phys.\ {\bf B253} (1985) 517.}
\itemitem{3.)}{I.J.R.\ Aitchison and C.M.\ Fraser, Ann.\ Phys.\ {\bf 156}
(1984) 1; D.\ Johnston, Nucl.\ Phys.\ {\bf B253} (1985) 687.}
\itemitem{4.)}{ N. Nakanishi, Prog. Theor. Phys. {\bf 35} (1966) 1111;
B. Lautrup, Mat. Fys. Medd. Dan. Vid. Selsk. {\bf 35} (1967) 29.}
\itemitem{5.)}{See e.g., C.\ Itzykson \& J.-B.\ Zuber, {\sl Quantum Field
Theory} (McGraw-Hill, New York, 1980).}
\itemitem{6.)}{P.A.M.~Dirac, \lqq Principles of Quantum Mechanics\rqq,
(OUP, Oxford, 1958) p.~302.}
\itemitem{7.)}{M.~Lavelle and D.~McMullan, Phys.~Lett.~{\bf 312B} (1993) 211.}
\itemitem{8.)}{S. Midorikawa, Prog. Theor. Phys. {\bf 61} (1979) 315.}
\itemitem{9.)}{D.\ Johnston, LPTHE Orsay Preprint 86/49 (unpublished).}
\itemitem{10.)}{P. Pascual \& R. Tarrach, {\sl QCD Renormalisation for the
Practitioner}  (Springer-Verlag, Berlin, 1984).}
\itemitem{11.)}{D.J.\ Broadhurst, N.\ Gray, K.\ Schilcher, Z. Phys.\ {\bf C52}
(1991) 111.}
\itemitem{12.)}{N.\ Gray, D.J.\ Broadhurst, W.\ Grafe, K.\ Schilcher, Z.\
Phys.\ {\bf C48} (1990) 673.}
\itemitem{13.)}{R.\ Tarrach, Nucl.\ Phys.\ {\bf B183} (1981) 384.}
\itemitem{14.)}{V.N.~Gribov, Nucl.~Phys.~{\bf B139} (1978) 1.}
\itemitem{15.)}{M.~Lavelle and D.~McMullan, \lqq Perturbative and
Non-Perturbative Quarks and Gluons\rqq, to appear in Physics Letters B.}
\itemitem{16.)}{V. Elias, T.G. Steele, M.D.~Scadron, Phys. Rev.
{\bf D38} (1988) 1584.}
\itemitem{17.)}{M.~Lavelle and M.~Oleszczuk,
Mod. Phys. Lett. {\bf A7} (Brief Reviews) (1992) 3617.}

\vfill\eject
\ni {\bf Figure Captions}

\ni {\bf Figure 1.} Radiative corrections to the mixed propagator
$\langle O\vert T\left(B(x) A_\mu(y)\right)\vert O\rangle$.  The dotted line
represents the field $B$.

\ni {\bf Figure 2.}  One-loop contributions to the Green function
${\cal F}_{\mu\nu}(p,-p,0)$.  The representation of $A_\mu$ and the
ghosts is conventional,
while the dotted line represents the field $B$.

\ni {\bf Figure 3.} One-loop contributions to the Green function
$\frac{p^\mu}{p^2}G^\mu(p,-p,0)$.  Diagram {\sl c} leads to massless
tadpoles and is thus zero
in dimensional regularization.

\ni {\bf Figure 4.} One-loop contributions to the Green function
$G^{(1)}(p,-p,0)$.
Diagram {\sl b} leads to massless tadpoles and is thus zero in dimensional
regularization.

\bye